\documentclass[preprint2]{aastex}

\usepackage{epsfig}
\usepackage{psfig}
\usepackage{lscape}

\newcommand{\HESS}{HESS~J1745-303}
\newcommand{\Ga}{G359.1-0.5}
\newcommand{\Gb}{G359.0-0.9}
\newcommand{\mouse}{PSR~J1747-2958}
\newcommand{\psr}{PSR~B1742-30}
\newcommand{\EG}{3EG~J1744-3011}

\slugcomment{}

\shorttitle{Fermi LAT observation of \HESS}
\shortauthors{Hui et al.}

\begin{document}

\title{Exploring the dark accelerator \HESS\ with \emph{Fermi} Large Area Telescope}

\author{C. Y. Hui\altaffilmark{1}, E. M. H. Wu\altaffilmark{2}, J. H. K. Wu\altaffilmark{2}, 
R. H. H. Huang\altaffilmark{3}, K. S. Cheng\altaffilmark{2}, 
P. H. T. Tam\altaffilmark{3}, and A. K. H. Kong\altaffilmark{3,4}}

\altaffiltext{1}{Department of Astronomy and Space Science, Chungnam
National University, Daejeon 305-764, Korea}
\altaffiltext{2}
{Department of Physics, University of Hong Kong, Pokfulam Road, Hong
Kong}
\altaffiltext{3}
{Institute of Astronomy and Department of Physics, National Tsing Hua University, Hsinchu, Taiwan}
\altaffiltext{4}
{Golden Jade Fellow of Kenda Foundation, Taiwan}


\begin{abstract}
We present a detailed analysis of the $\gamma-$ray emission from \HESS\ with the data obtained by the \emph{Fermi} 
Gamma-ray Space Telescope in the first $\sim29$ months observation. 
The source can be clearly detected at the level of $\sim18\sigma$ and $\sim6\sigma$ in 
$1-20$~GeV and $10-20$~GeV respectively. Different from the results obtained by the \emph{Compton} Gamma-ray Observatory, 
we do not find any evidence of variability. Most of emission in $10-20$~GeV is found to coincide with the region C
of \HESS. A simple power-law is sufficient to describe the GeV spectrum with a 
photon index of $\Gamma\sim2.6$. The power-law spectrum inferred in the GeV regime can be connected to that of a particular spatial 
component of \HESS\ in $1-10$~TeV without any spectral break. These properties impose independent constraints for 
understanding the nature of this ``dark particle accelerator". 
\end{abstract}


\section{INTRODUCTION}

Recent surveys of the central region of our Galaxy with the {\bf H}igh {\bf E}nergy {\bf S}teroscopic {\bf S}ystem
(H.E.S.S.) have uncovered a number of $\gamma-$ray sources in the TeV regime (Aharonian et al. 2002, 
2005a, 2005b, 2006a). Different from the cases of pulsar wind nebulae (PWNe) and young supernova 
remnants (SNRs), some of these sources have no non-thermal X-ray counterpart yet been identified. 
Among them, \HESS\ is one of the most enigmatic objects. 

\HESS\ was firstly discovered by the H.E.S.S. Galactic Plane Survey (Aharonian et al. 2006a) and was 
subsequently investigated in details with dedicated follow-up observations (Aharonian et al. 2008). 
The TeV $\gamma-$ray image shows that it consists of three spatial components (i.e. Regions A, B and C 
in Fig.~1 of Aharonian et al. 2008). Owing to the lack of spectral variability and the insignificant 
dip among these regions in the existing data, it was argued that they are originated from a single 
object (Aharonian et al. 2008). This inference suggests that \HESS\ as one of the largest 
unidentified TeV sources which has an angular size of $\sim0.3^{\circ}\times0.5^{\circ}$. 

Searches for the possible non-thermal diffuse X-ray component in region A of \HESS\ 
have been conveyed with \emph{XMM-Newton} and \emph{Suzaku} (Aharonian et al. 2008; 
Bamba et al. 2009). None of these observations have resulted in any evidence for the diffuse X-ray 
emission. This imposes a TeV-to-X-ray flux ratio larger than $\sim4$ (Bamba et al. 2009) which 
is larger than the typical value of PWNe and SNRs (i.e. less than 2) 
(cf. Matsumoto et al. 2007; Bamba et al. 2007, 2009). 
Because of the non-detection 
of counterparts in X-ray/radio, \HESS\ is dubbed as a ``dark accelerator" (Bamba et al. 2009).

While no non-thermal diffuse X-ray emission has yet been found, a possible excess of 
neutral iron line emission was discovered in the direction toward the region A of \HESS\ 
(Bamba et al. 2009). Together with its proximity to the Galactic center and the positional conicidence 
of a molecular cloud (Aharonian et al. 2008), the line emission is suggested to be the reflected 
X-rays originated from the previous activity in the Galactic center (Bamba et al. 2009). This molecular
cloud can be interacted with the shock from a nearby SNR \Ga\ (Bamba et al. 2000, 2009; Lazendic et al. 2002; 
Ohnishi et al. 2011) and 
produce the observed $\gamma-$rays through the acceleration of protons and/or leptons (see Bamba et al. 2009). 
Nevertheless, this proposed scenario cannot be confirmed unambiguously. In view of the presence 
of many surrounding objects (see Fig.~1 in Aharonian et al. 2008), including the ``mouse" pulsar 
(i.e. \mouse), one cannot rule out these objects as the source of energetic particles simply based on the TeV 
results (Aharonian et al. 2008). Furthermore, with the current information, it is not possible to 
discriminate hadronic model and leptonic model (see Fig.~5 \& 6 in Bamba et al. 2009). 
In order to do so, investigations in lower energy regime are required. 

It is interesting to note that \HESS\ is positionally coincident with an unidentified EGRET source 
\EG\ (Hartman et al. 1999). Different from \HESS, \EG\ was suggested to demonstrate long-term 
variability (Torres et al. 2001). Also, based on the MeV-GeV spectrum observed by EGRET, 
the extrapolated flux in the TeV regime overshoots that observed by H.E.S.S.. 
Therefore, Aharonian et al. (2008) considered that \EG\ is unrelated to \HESS. 

After the commence of the Large Area Telescope (LAT) onboard 
\emph{Fermi} Gamma-ray Space Telescope, a detailed investigation of this dark accelerator 
in the MeV$-$GeV regime is now feasible with its much improved spatial resolution and sensitivity. 
However, among 1451 objects detected by LAT during the first 11 months, we do not identify any source
corresponding to \HESS\ / \EG\ (Abdo et al. 2010). Very recently, in an analysis of the $\gamma-$rays 
from the Galactic center with first 25 months LAT data, a new serendipitous source was found to 
coincide spatially with \HESS\ / \EG\ (Chernyakova et al. 2011). 
In this paper, we report a detailed analysis of this source with LAT observation in the first 
$\sim29$~months. 

\section{DATA ANALYSIS \& RESULTS}

In this analysis, we used the data obtained by LAT between 2008 August 4 and 2010 December 23. The 
\emph{Fermi} Science Tools v9r18p6 package is used to reduce and analyze the data in the vicinity of \HESS.  
Only the events that are classified as class~3 or class~4 are adopted. The post-launch
instrument response functions ``P6\_V3\_DIFFUSE" were used throughout this investigation. 

With the aid of the task \emph{gtlike}, we performed unbinned maximum-likelihood analysis for 
a circular region-of-interest (ROI) with a $10^{\circ}$ diameter centered on the nominal position of \HESS\ 
(i.e. RA=$17^{h}$$45^{m}$$1.999^{s}$ Dec=$-30^{\circ}$$22^{'}$$12.0^{''}$ (J2000)). 
The size of ROI have been chosen to avoid the surrounding bright sources so as 
to reduce the systematic uncertainties due to the inaccurate background subtraction in this complex region. 
For subtracting the background contribution, we included the Galactic diffuse model (gll\_iem\_v02.fit) and the 
isotropic background (isotropic\_iem\_v02.txt), as well as 41 sources in the first \emph{Fermi}/LAT 
catalog (1FGL; Abdo et al. 2010) within $10^{\circ}$ from the aforementioned center. 

To begin with, we compared the spectral properties inferred by LAT and EGRET. For a consistent comparison with 
Hartman et al. (1999), we used events with energies $>100$~MeV for our initial analysis. 
We assumed a power-law (PL) spectrum for \HESS\ as well as all 1FGL sources in our consideration. 
All the sources are assumed to be point sources throughout this investigation. 
The best-fit model yields a photon index of $\Gamma=2.16\pm0.03$\footnote{All errors quoted in 
this paper are statistical only and are computed for a confidence interval of $1\sigma$.} 
and a test-statistic (TS) value of 499 which corresponds to a 
significance of $22\sigma$. This is consistent with the significance reported in the preliminary analysis by 
Chernyakova et al. (2011). The photon index is found to be consistent with \EG\  
(i.e. $\Gamma=2.17\pm0.08$). To further compare with the EGRET results, we compute the integrated photon flux 
in $0.1-10$~GeV which is found to be $2.06^{+0.24}_{-0.23}\times10^{-7}$~cm$^{-2}$~s$^{-1}$. This 
is $\sim3$ times smaller than that of \EG\ (i.e. $6.39\pm0.71\times10^{-7}$~cm$^{-2}$~s$^{-1}$). The discrepancy can be due 
to the improved spatial resolution of LAT, and hence the estimation of background contribution from the nearby sources 
is more accurate than EGRET.

\begin{figure}[t]
\centerline{\psfig{figure=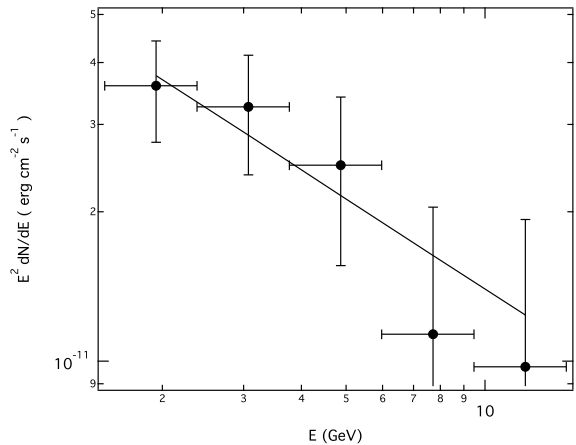,width=8cm,clip=,angle=0}}
\caption[]{\emph{Fermi} LAT spectrum of \HESS. The solid line represents the best-fit power-law model.} 
\label{spec}
\end{figure}

Considering the background, as \HESS\ is located close to the Galactic center where the diffuse $\gamma-$ray 
emission is very intense, the contamination can possibly be large at lower energies. 
Also, the point spread function (PSF) of LAT is narrower at higher energies. 
While the $68\%$ containment radius at 100~MeV is $\sim4.5^{\circ}$, it is only $\sim0.8^{\circ}$ at 1~GeV.
\footnote{For updated status, please refer to http://www-glast.slac.stanford.edu/software/IS/glast\_lat\_performance.htm}
Therefore, the contamination due to 
the PSF wings of the nearby sources can also be minimized by limiting the analysis at higher energies. 
In order to minimize the systematic uncertainty in the background modeling so as to obtain robust results, we 
restricted the subsequent analysis in $1-20$~GeV. In this adopted band, the best-fit PL model results in a TS value of 332
which corresponds to a detection significance to be $\sim18\sigma$. To examine the robustness of the detection, 
we have also considered the systematic uncertainty of the Galactic diffuse emission background. Following Abdo et al. (2010), we 
repeat the analysis by varying the slope and the normalization of the Galactic diffuse model in $0-0.07$ and $\pm10\%$ respectively.
Within the uncertainty of the background model, the detection significance remains over $10\sigma$.
This best-fit PL model yields a photon index of $\Gamma=2.60\pm0.05$ and an energy flux of 
$5.25^{+1.87}_{-1.47}\times10^{-11}$~ergs~cm$^{-2}$~s$^{-1}$ in $1-20$~GeV. 
The best-fit PL model and the $\gamma-$ray spectrum as seen by \emph{Fermi} LAT is shown in Figure~\ref{spec}.

Besides a simple PL model, we have also examined if a broken power-law (BKPL) or an 
exponential cutoff power-law (PLE) can describe the spectrum better. 
The fittings with BKPL and PLE results in the TS values of 335 and 333 respectively. 
Based on the likelihood ratio test, the additional spectral parameters in BKPL/PLE are not strongly required
which suggests that a single PL is statistically sufficient to describe the data.
For the PLE model, the best-fit photon index and cut-off energy are $\Gamma=2.31\pm0.07$ and
$E_{\rm cutoff}=12.63\pm4.69$~GeV respectively. On the other hand, for the BKPL, the initial fitting resulted 
in a break energy and the photon indicies of $E_{\rm break}=3.68\pm0.22$~GeV, $\Gamma_{1}=2.32\pm0.15$ and $\Gamma_{2}=3.09\pm0.26$ 
respectively. However, different from the best-fit solutions inferred from PL and PLE models, we found that 
the solution inferred from the BKPL fit is unstable subjecting to the perturbations in the parameter space. 
In view of the problem of the convergence, we will not further consider BKPL model in all subsequent analysis. 

Although the extra parameter, $E_{\rm cutoff}$, in the PLE model is not statistically required, one is not able to completely rule 
out it in view of its capability in depicting the observed data. As a PLE model typically describes the $\gamma-$ray spectrum of a pulsar (Abdo et al. 2010b, 2011), 
we speculate if there is any hidden pulsar in \HESS. To test this hypothesis, we have performed a blind search for 
coherent pulsation. The arrival times of each photon were barycentric corrected with the nominal position of \HESS\ adopted. 
To minimize the impact due to the ignorance of the spin-down rate, we have divided the full 
time span of the adopted data into 5 segments and run the Fourier analysis 
in each segment independently with the aid of the tool \emph{gtpspec}. From each computed power spectrum, we picked out 
10 peaks and investigated if there is any correlation among different segments. However, we did not 
identify any promising periodicity candidate in this data. Hence, we conclude that there is no evidence for a hidden pulsar in \HESS. 

\begin{figure*}[t]
\centerline{\psfig{figure=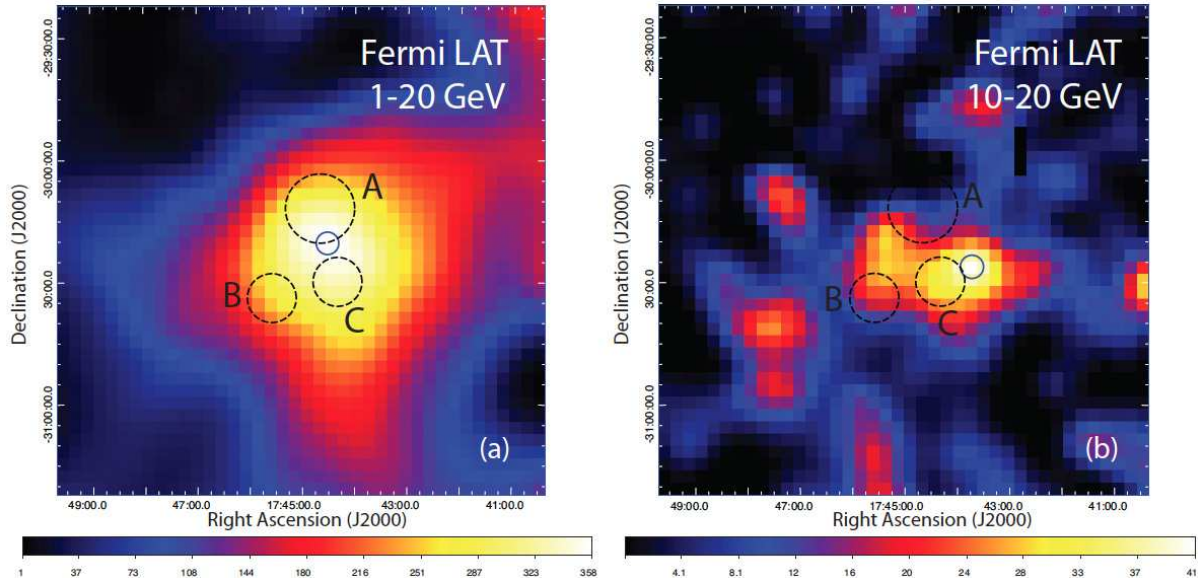,width=16cm,clip=,angle=0}}
\caption[]{{\bf (a)} Test-statistic (TS) map in $1-20$~GeV of a region of $2^{\circ}\times2^{\circ}$ 
centered at the nominal position 
of \HESS. The color scale that used to indicate the TS value is shown by the scale bar below. The blue circle 
represents the $1\sigma$ positional error circle determined by \emph{gtfindsrc}. Various TeV emission 
components of \HESS\ (i.e. regions A, B and C in Fig.~1 of Aharonian et al. 2008) are illustrated by the 
black dashed circles. 
{\bf (b)} Same as Fig.~2a but in the energy range of $10-20$~GeV.}
\label{ts}
\end{figure*}

As \EG\ was reported to be a variable (Torres et al. 2001), we also examine the  
variability with LAT data. First, we extracted the 
light curve obtained from the data within $1^{\circ}$ from the nominal position of \HESS\ with a binning 
factor of 10 days in the energy range of $1-20$~GeV. By fitting a horizontal line to the light 
curve, we obtain a reduced chi-square of $\chi^{2}_{\nu}=1.5$ for 87 degrees of freedom. 
Hence, there is no strong evidence for any variability or flaring. 

For a further investigation of the possible spectral and flux variability, we divided the whole data into 
five segments of equal time span and performed an unbinned likelihood analysis on each segment. The 
results are summarized in Table~\ref{spec_var}. A simple PL was adopted for all the fittings. Within the tolerance 
of the statistical uncertainties, we conclude that neither the spectral shape nor the flux varies among these 
segments.

\begin{center}
\begin{deluxetable}{ccc}
\tablewidth{0pc}
\tablecaption{$\gamma-$ray spectral properties and energy fluxes of \HESS\ at different epochs.}
\startdata
\hline\hline
 Time segment & $\Gamma$ & $f_{\gamma}$\tablenotemark{a} ($1-20$~GeV) \\
   Mission elapsed time  (s)     & & $10^{-11}$~ergs~cm$^{-2}$~s$^{-1}$ \\\hline
239557417$-$254601478.2   & $2.67\pm0.23$ & $4.45^{+12.93}_{-3.96}$ \\
254601478.2$-$269645539.4 & $2.45\pm0.18$ & $6.65^{+14.28}_{-5.36}$ \\
269645539.4$-$284689600.6 & $2.57\pm0.05$ & $5.73^{+2.22}_{-1.7}$ \\
284689600.6$-$299733661.8 & $2.70\pm0.25$ & $5.02^{+16.15}_{-4.60}$ \\
299733661.8$-$314777723   & $2.80\pm0.30$ & $4.35^{+19.16}_{-4.28}$ \\
\enddata
\label{spec_var}
\tablenotetext{a}{The quoted errors of energy have taken the statistical uncertainties of both photon index and prefactor
into account.}
\end{deluxetable}
\end{center}

We have computed the $2^{\circ}\times2^{\circ}$ TS map in $1-20$~GeV centered at the nominal position of \HESS\ by
using the tool \emph{gttsmap}. This is displayed in Figure~\ref{ts}a. Utilizing \emph{gtfindsrc}, we  
determined the best-fit postion in $1-20$~GeV to be 
RA=$17^{h}$$44^{m}$$31.440^{s}$ Dec=$-30^{\circ}$$20^{'}$$32.28^{''}$ (J2000)
with a $1\sigma$ error radius of $0.05^{\circ}$. The position locates between regions A and C of 
\HESS. Given that the PSF has a 68\% containment radius of $\sim0.8^{\circ}$ at 1~GeV, 
the source extent inferred in this band is consistent with that of a point source. 
The relatively wide PSF do not allow us to determine whether GeV emission is associated with any particular TeV feature. 
To further examine the spatial nature, we also computed the TS map in $10-20$~GeV which is displayed in 
Figure~\ref{ts}b. Since the 68\% containment radius at 10~GeV is $\sim4$ times smaller than that at 1~GeV, 
the feature can be better resolved. We found that the peak TS value found in this band is 41 which corresponds 
to a significance of $\sim6\sigma$. Within the systematic uncertainty of the Galactic 
diffuse background, we found that the detection significance of the source remains over $4\sigma$ in this band.
The best-fit position in $10-20$~GeV is found to be
RA=$17^{h}$$43^{m}$$44.160^{s}$ Dec=$-30^{\circ}$$26^{'}$$24.00^{''}$ (J2000)
with a $1\sigma$ error radius of $0.05^{\circ}$. This differs from that inferred in $1-20$~GeV by $0.2^{\circ}$. 
We note that the GeV feature found in this hard band is apparently peaked at region C and possibly extended to region B. 
Although it appears to be extended in the hard band, the relatively low detection significance in this energy range 
does not allow a firm conclusion.

\section{DISCUSSION}

In this paper, we have reported a detailed study of \HESS\ with \emph{Fermi} LAT data which provides a missing 
piece in understanding the nature of this dark accelerator. In view of the putative variability of \EG, 
Aharonian et al. (2008) argued that the GeV counterpart is unlikely to be associated with \HESS. However, we do not 
find any evidence for the spectral/flux variability from the LAT data. The discrepancy of the LAT result and that inferred 
from EGRET is most likely due to the differences between their instrumental performance. With the improved angular 
resolution and sensitivity of LAT, many previously unknown sources in the proximity of \HESS\ are now detected. 
As several sources within $\sim2^{\circ}$ are found to be variable, including 1FGL~J1747.2-2958 which is the $\gamma-$ray 
counterpart of \mouse\ (Abdo et al. 2010), the source confusion in the EGRET data can possibly lead to the apparent 
variability. 

It is interesting to compare the spectral properties inferred in the GeV regime with those obtained 
by H.E.S.S.. First, the photon index inferred from the LAT data (i.e. $\Gamma=2.60\pm0.05$) is similar to that inferred in 
TeV (cf. Tab.~2 of Aharonian et al. 2008). Furthermore, the extrapolated photon flux in $1-10$~TeV with the best-fit 
PL model is found to be $(2.21^{+1.84}_{-1.03})\times10^{-13}$~cm$^{-2}$~s$^{-1}$ which is 
consistent with any individual spatial component observed by H.E.S.S. within $1\sigma$ uncertainties 
(cf. Tab.~2 of Aharonian et al. 2008). 

\begin{figure}[t]
\centerline{\psfig{figure=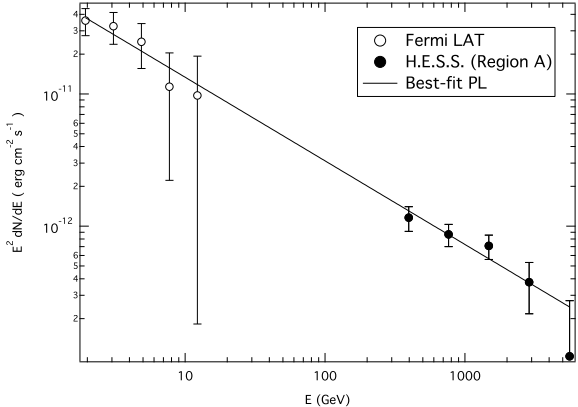,width=8cm,clip=,angle=0}}
\caption[]{Spectral energy distribution of \HESS\ as observed by \emph{Fermi} LAT and H.E.S.S. The H.E.S.S. spectrum is for region A. 
The solid line represents the best-fit power-law model inferred in the joint analysis of both data sets.}
\label{sed}
\end{figure}

Since the TeV spectral data of region A is avaliable (cf. Fig.~2 of Aharonian et al. 2008)
\footnote{http://www.mpi-hd.mpg.de/hfm/HESS}, 
we further compare this particular region with the GeV spectrum by constructing a spectral 
energy distribution (SED) which is display in Figure~\ref{sed}. Both data can be simultaneously fitted with 
a single PL of $\Gamma=2.63\pm0.03$. This clearly demonstrates that the TeV spectrum of this spatial component 
can be smoothly connected to the GeV spectrum. 

For regions B and C, despite that no TeV spectral data is currently available, we note that their spectral shapes 
reported by Aharonian et al. (2008) are similar to that of region A. Within the statistical uncertainties of the 
spectral properties inferred in both GeV and TeV regimes, the TeV spectra of these regions can also be possibly 
connected to that of GeV. Further investigation by H.E.S.S. is strongly encouraged, particularly for region C as 
most of the $\gamma-$ray emission in $10-20$~GeV apparently coincides with this spatial component. 


As both PL and PLE model can describe the GeV spectrum equally well, we cannot discriminate these competing models 
based on the current data. In view of the exponential cut-off, the spectral connection between the GeV and TeV regimes 
cannot be established with the PLE model. However, we would like to point out that the best-fit cut-off energy, 
$E_{\rm cutoff}=12.63\pm4.69$~GeV, falls in the highest energy bin of the LAT spectrum (cf. Fig.~\ref{spec} and Fig.~\ref{sed}). 
Owing to the small photon statistics in the hard band, the statistical uncertainty of the highest energy bin is rather large. 
In view of this, it remains unclear whether the inferred cut-off is genuine as this particular data point is sensitive to the 
systematic uncertainty of the background. 
Analysis of the LAT data in higher energies (e.g. $\gtrsim10$~GeV) with sufficient 
photon statistic in the future can help to discriminate these two models. 

The detection of $\gamma-$rays provides a strong evidence for the particle acceleration. It should 
be noted that there are two SNRs and two pulsars around \HESS\ which can be the potential high energy particle injector 
(cf. Fig.~1 of Aharonian et al. 2008). Based on the small distances of \Gb\ and \psr, Bamba et al. (2009) argued that these two 
sources are foreground objects which are unrelated to \HESS. With the GeV counterpart revealed by LAT, 
we can now safely rule out the possibility that \psr\ is the major contributor as its spin-down flux, 
$\dot{E}/4\pi d^{2}\sim10^{-11}$~erg~cm$^{-2}$~s$^{-1}$, is lower than the energy flux observed in the GeV regime 
(cf. Manchester et al. 2005). On the other hand, for \mouse, the sum of $\gamma-$ray flux of 1FGL~J1747.2-2958 
and \HESS\ observed by LAT only consumes $\sim5\%$ of its spin-down flux ($\sim3\times10^{-9}$~erg~cm$^{-2}$~s$^{-1}$). 
Therefore, it is energetically possible to be the source for the high energy particles. 
Nevertheless, with its eastward proper-motion
(Hales et al. 2009), its backward extrapolated position by its spin-down age is found to have 
an offset of $\sim0.6^{\circ}$ and $\sim0.8^{\circ}$ from the best-fit positions inferred in $1-20$~GeV and 
$10-20$~GeV respectively. 
If the feature found by H.E.S.S. and LAT is indeed a PWN, then it is one of 
the most peculiar system because of its large positional offset with respect to the pulsar position. 

While the contribution of \mouse\ is uncertain, the interaction between the shock from \Ga\ and a molecular cloud in its 
neighbourhood is considered to be a more viable means to produce the observed $\gamma-$rays. For both hadronic and leptonic scenarios, 
Bamba et al. (2009) have modeled the broadband spectrum of the region A of \HESS. However, comparing Fig.~\ref{spec}
in this paper and Fig.~5 \& 6 in Bamba et al. (2009), 
the GeV flux predicted in all the scenarios considered in their work are at least an order of magnitude lower 
than that observed by LAT. Revised theoretical investigation with constraints provided by LAT 
is therefore required. Moreover, while the other known systems of SNRs interacting
with molecular clouds have their remnant shells found in other wavelengths 
coincided with the $\gamma-$ray emission (e.g. Castro \& Slane 2010; Abdo et al. 2010c), 
there is no such evidence for \HESS. We should point out that
all the previous multiwavelength campaign were targeted only at the bright component of \HESS, namely the region A. 
Since the LAT observation suggested that the GeV emission can possibly be originated from the regions C/B, observational 
investigations aim at these regions are encouraged for a further understanding of this mysterious object. 

Based on the above discussion, we have to admit that the energy injection source of \HESS\ remains unclear. It is instructive to 
compare it with other nearby high energy sources. Aharonian et al. (2006b) have reported observations of an extended region of 
very high energy (VHE, $>100$~GeV) $\gamma$-ray emission correlated spatially with a complex of giant molecular clouds in the central 
200~pc of the Milky Way. It appears that TeV emissions from the molecular clouds in the vicinity of the Galactic Center are quite 
common phenomena. In addition, similar to the case of \HESS, 6.4 keV lines have been commonly detected from many molecular clouds near 
the Galactic Center, e.g. Sgr~B2 (Koyama et al. 1996; Murakami et al. 2000), Sgr~C (Murakami et al. 2001) and others (Bamba et al. 2002; 
Predehl et al. 2003; Nobukawa et al. 2008; Nakajima et al. 2009; Koyama et al. 2009). It has been speculated that these 
neutral iron lines arise from the reflection by the dense molecular clouds which are irradiated by the nearby X-ray sources 
(e.g. Koyama et al. 2007; Inui et al. 2009). On the other hand, Dogiel et al. (2009) argued that these 
6.4~keV lines from the molecular clouds are excited by a background intensity of subrelativistic protons coming from the escaped 
part of a past captured star by the Galactic supermassive black hole Sgr~A*. The periodic stellar capture 
events may explain the recent observed {\it Fermi} bubble (Cheng et al. 2011a). 
This subrelativistic proton wind can also form shock by hitting 
the clouds and produce relativistic protons. It is shown that the decay of neutral pions produced by hadronic collisions between 
the accelerated relativistic protons in the clouds can emit a power-law $\gamma-$ray spectrum 
from 30 MeV to 10 TeV without any spectral break (Cheng et al. 2011b in preparation). 
If this is true, then the high energy emission from various molecular clouds in the vicinity of the Galactic center are correlated to the 
past activities of the Galactic center and their intensity might be correlated with the propagation history of the injection of these 
subrelativistic protons escaped from the capture past events.

\acknowledgments{
CYH is supported by research fund of Chungnam National
University in 2010.
KSC is supported by a GRF grant of Hong Kong Government
under HKU 7011/10P.
And AKHK is supported partly by the National
Science Council of the Republic of China (Taiwan)
through grant NSC99-2112-M-007-004-MY3 and a Kenda
Foundation Golden Jade Fellowship.}


\end{document}